\documentclass{JHEP3} 

\usepackage{epsfig,multicol}
\usepackage{amsmath, amssymb}
\usepackage{epic}
\usepackage{concmath, palatino}

\newcommand\fverb{\setbox\pippobox=\hbox\bgroup\verb}
\newcommand\fverbdo{\egroup\medskip\noindent%
			\fbox{\unhbox\pippobox}\ }
\newcommand\fverbit{\egroup\item[\fbox{\unhbox\pippobox}]}
\newbox\pippobox

\newcommand{\sxx}{\nonumber \\ && }

\newcommand{\E}{{\cal E}}
\newcommand{\Sp}{{\cal S}}
\newcommand{\J}{{\cal J}}

\newcommand{\be}{\begin{equation}}
\newcommand{\ee}{\end{equation}}
\newcommand{\ba}{\begin{eqnarray}}
\newcommand{\ea}{\end{eqnarray}}

\newcommand{\refeq}[1]{Eq.~(\ref{eq:#1})}

\newcommand{\ads}{AdS_5\times S^5}

\def\bi{\bibitem}

\allowdisplaybreaks

\title{Reciprocity and integrability in the 
$\mathfrak{sl}(2)$ sector of ${\cal N}=4$ SYM}

\author{Matteo Beccaria and Guido Macorini\\
  Dipartimento di Fisica, Universita' del Salento, 
  Via Arnesano, 73100 Lecce\\
  INFN, Sezione di Lecce\\
  E-mail: \email{matteo.beccaria@le.infn.it}, \email{guido.macorini@le.infn.it} 
}

\abstract{
We analyze the higher conserved charges of type IIB superstring on $\ads$ from the perspective of 
a recently discovered generalized Gribov-Lipatov reciprocity. We provide several evidences that 
reciprocity holds for all the higher charges and not only for the energy. This is discussed in the 
simple case of twist $L=2$, and $3$ operators in the $\mathfrak{sl}(2)$ subsector at (a) multi-loop level 
in weak coupling, (b) classical level at strong coupling for the dual folded string.     
}

\begin{document} 

\section{Introduction and Discussion}

In the last years the discovery of integrable structures in the contest of
the AdS/CFT correspondence~\cite{Maldacena:1997re} has led to an impressive amount  of new results, 
opening indeed a new direction in the search of quantitative test of the
conjecture~(see for example the recent review~\cite{Tseytlin:2009zz}).\\
 
The first fundamental step in this direction was carried out in the seminal
paper~\cite{Minahan:2002ve}, where it was realized that the dilatation
operator in the in the  $\mathfrak{so}(6)$ subsector of the gauge theory, 
in the large $N$ planar limit,  acts on composite single trace
gauge invariant operators as an Hamiltonian of a generalized spin chain; 
this approach was soon extended to the whole theory~\cite{Beisert:2003jj,Beisert:2003yb,Beisert:2005fw}.\\

After the translation to the spin chain formalism, one can employ the powerful algebraic
techniques of the Bethe Ansatz. Using this approach 
a number of impressive multi-loops results has been
obtained~\cite{Beccaria:2009eq,Kotikov:2007cy,Bajnok:2009vm,Christoph}, 
while the corresponding direct calculations 
using the standard field theory techniques would have been (almost) hopeless.\\

As usual, 
integrability emerges as factorized scattering of elementary excitations due to 
the existence of 
an infinite set of commuting conserved charges $q_k$.
On the string side of the correspondence, the classical $\sigma$-model 
describing the  strings in the curved background $\ads$ was proved to be 
a (classical) two dimensional integrable model; in fact, 
the model can be related to the integrable Neumann system~\cite{Arutyunov:2003uj}; the infinite set of nonlocal
classically conserved charges was found in~\cite{Bena:2003wd,Mandal:2002fs,Engquist:2004bx}.\\

Despite  the great interest in the integrability structures emerging in the
AdS/CFT correspondence, the properties of the higher conserved charges have
not been investigated at the same level of the first of them. Actually, $q_2$ represents the eigenvalues
of the dilatation operator, {\em i.e.} the anomalous scaling dimensions, or the energy of the dual 
string states. As such, it has a clear cut physical interpretation allowing. 
Although, the correspondence of the full tower of charges between weak and strong coupling 
has been investigated~\cite{Arutyunov:2003rg}, the physical properties of higher charges
remain, in our opinion, less clear. \\
 
The aim of this paper is to begin an investigation of such features working in the framework of the 
$\mathfrak{sl}(2)$ subsector. At weak coupling, this is a closed subsector under perturbative
renormalization, composed by (linear combinations of) single trace operators of
the form
\begin{equation}
{\rm Tr}\,(\mathcal{D}^{n_1} Z \dots \mathcal{D}^{n_L} Z)
\end{equation} 
where $N=\sum n_i$ is the total spin and the number of fields $L$ is called the twist
of the operator (the twist corresponds to the classical dimension minus the
spin). In particular we will focus on the recently proposed generalized 
Gribov-Lipatov reciprocity~\cite{dok1,bk,dok2}.\\
It is a property that arises in QCD where the crossed processes of deep
inelastic scattering  and $e^+e^-$ annihilation can be treated in a
symmetric approach based on modified DGLAP evolution equations for 
parton distributions. The modified DGLAP kernel $P(N)$ obeys perturbatively 
\be
\label{nonlinear}
\gamma(N) = P\left(N+\textstyle{\frac{1}{2}\gamma(N)}\right),
\ee
where $\gamma(N)$ is the lowest anomalous dimension. 
Then, the reciprocity constraints is simply the claim that the large spin 
$N$ behavior of $\gamma(N)$~\cite{bk} 
can be written as the following asymptotic condition
\be
\label{eq:RRJ}
P(N) = \sum_{\ell\ge 0} \frac{a_\ell(\log\,J^2)}{J^{2\,\ell}}, \quad J^2 = N\,(N+1),
\ee  
where $a_\ell$ are suitable coupling-dependent po\-ly\-nomials and $J^2$ is the 
Casimir of the col\-li\-near subgroup $SL(2, \mathbb{R})\subset SO(2,4)$ 
of the  conformal  group~\cite{bra}. \refeq{RRJ} can be read as parity 
invariance under (large) $J\to -J$.\\

An important point to be stressed is that reciprocity is not a rigorous
prediction, but a physically motivated property which 
requires explicit tests at higher loop order. It is known to hold 
in many QCD and $\mathcal{N}=4$ SYM multi-loop calculations~\cite{Forini:2008ky}. 
In particular, four-loop reciprocity of the twist 2 supermultiplet in 
$\mathcal{N}=4$ SYM has been proved in~\cite{spires1}. A similar result for the five-loop 
anomalous dimension of twist-3 operators can be found in~\cite{Beccaria:2009eq}. 
Other sectors have also been investigated, like 
the gluonic sector discussed in~\cite{bf}.\\



%


On the string side the dual partner of the gauge operators is identified with 
the so called folded string $(S,J)$ solution: geometrically it describes 
a string stretched along the radial direction of $AdS_5$ and rotating in
$AdS_5$, with center of mass moving on a circle of $S^5$~\cite{folded}. 
The first hint of the presence of reciprocity properties for the anomalous
dimensions in the strong regime was found in~\cite{Basso:2006nk} and the analysis was extended
in~\cite{Beccaria:2008tg}, where the property has found to hold also for
the leading string 1-loop corrections. 
The higher charges at strong coupling have been computed in~\cite{Arutyunov:2003rg} for a string configuration $(J_1, J_2)$ with two 
angular momenta on $S^5$ and related to the $(S,J)$ folded case by analytic continuation~(see the review \cite{Tseytlin:2003ii})

\begin{equation}
 E \to -J_1,\qquad S \to J_2,\qquad J \to  -E
\end{equation}

The analytic continuation relates the equations of motion and the conserved
charged of the two $\sigma$-models.\\ 


Since from the integrability point of view all the conserved charges are on
the same ground (i.e. the energy does not play a special role) it is natural to
ask whether the higher conserved charges share the same reciprocity behavior.
The result of this paper is indeed a substantial evidence that 
reciprocity holds for all the higher charges and not only for energy. \\

Our analysis will consider the minimal anomalous dimension non degenerate state in the class of twist 2 or 3 operators. 
The reason for such constraint comes from the following leading order (one-loop in weak couping and classical level in string theory) 
discussion~\footnote{We thank G. Korchemsky for many 
very helpful discussions on this point.}. Parity invariance
implies analytical continuation of the energy spectrum in the spin $N$
which is an extremely nontrivial issue. Instead, we can consider the limit
of large $N$ where we can obtain asymptotic expansion of the energy
and, then, discuss the properties of the asymptotic series under parity.
Quantized values of the energy and conserved charges form trajectories~\cite{Belitsky:2004cz} which are 
enumerated by integers $n_k$.
It is therefore quite natural that reciprocity/parity relation should
also act on these integers. Besides, all energy levels except
the minimal one are double degenerate. As a consequence, the minimal
energy trajectory should go into itself under $N\to -N-1$ while excited
trajectories could mix with each other. Namely, denoting the coefficients of the transfer matrix $\widetilde{q}_k$ (closely related to 
the conserved charges), the corresponding transformation reads
$\widetilde{q}_k \to (-1)^k \widetilde{q}_k$.
This transformation leaves invariant the spectral
curve of the $\mathfrak{sl}(2)$ spin chain and the same arguments apply to the spectral
curve of $\ads$ sigma model. Therefore, the following consideration
applies both in gauge and in string theory. \\

Moreover, as discussed in~\cite{Korchemsky:1996kh}, the quantization condition for the conserved
charges follow from the requirement for periods of the action to
take values parameterized by the integers $n_k$ mentioned above. 
To study reciprocity/parity, the quantization conditions should
be expressed in terms of $-q_2$. However, the quantized values of the integrals of motion 
turns out to depend also on the moduli $\delta_k = (n_k+1/2)/(-q_2)^{1/2}$. If these
moduli were absent, the integrals of motion were parity invariant (like
the minimal energy trajectory). The presence of the moduli makes things
more involved. The cases $L=2, 3$ are precisely free from this complications.\\

In the weak coupling regime we computed the first two non vanishing charges
$q_{4,6}$ at three-loops (plus the four-loops dressing part) and two-loops
respectively, while on the strong coupling side we tested the first ten
charges at classical level.
Our result is that the kernel $P_r$ appearing in the relation 
\be
q_r(N) = P_r(N + \frac{1}{2} q_2(N)),
\ee
is reciprocity respecting in the sense of \refeq{RRJ}. The change
from the bare conformal spin $N$ to the
renormalized $N + \frac{1}{2}\,q_2(N)$ must be done for the full tower of conserved charges. The
 fact that higher charges are functions of
$N+ \frac{1}{2} q_2 = N + \frac{1}{2} ( E-N) = \frac{1}{2}( E + N)$
is natural from the point of view of light-cone quantization~\cite{spires3} where 
everything depends on $N$ via the momentum component $p^+$~\footnote{We thank A. A. Tseytlin for this 
important comment.}. At strong coupling, a similar relation holds with $N$ and $q_2(N)$ being replaced by the scaled spin ${\cal S} = N/\sqrt\lambda$ 
and energy ${\cal E} = E/\sqrt\lambda$ which are kept fixed in the semiclassical limit as the coupling $\lambda\to \infty$. 
A set of independent quantities playing the role of the weak coupling charges $q_r$ is identified at strong coupling exploiting
the abovementioned analytical continuation from the $(J_1, J_2)$ string.\\

To conclude, we have expanded the scope of reciprocity in $\ads$, including the natural discussion of higher charges. 
Of course, a few words are deserved to the interplay between reciprocity and finite size wrapping effects, {\em i.e.} TBA investigations.
From this point of view, we have to tell between the weak coupling and strong coupling regimes. In the former, 
reciprocity has been established to be respected by wrapping corrections. Besides, it can be used as an efficient tool in order to constrain
analytical prediction. A paramount example is the five loop computation of twist-2 anomalous dimension which includes a NLO wrapping correction~\footnote{A. Rej, private communication}.
Wrapping corrections to higher charges have not yet been studied, but are hardly expected to violate reciprocity, at least in our opinion.
At strong coupling, things are less clear since one needs at least a one-loop analysis of the semiclassical string energies as well as an accurate 
expansion of energy or other conserved quantities at large spin.


%
%


\section{Analytic computation of $q_r$ from the Baxter equation}
\label{sec:analytic}

We follow the BES convention~\cite{Beisert:2006ez} for the definition 
of the weak coupling expansion parameter
\be
g = \frac{\sqrt\lambda}{4\,\pi},
\ee
and the $x$-variables are defined as
\be
x(u) = \frac{u}{2}\,\left(1+\sqrt{1-\frac{4\,g^2}{u^2}}\right).
\ee
The conserved charges are (the usual anomalous dimension is $q_2$)
\be
q_r(M, g) = 2\,i\,g^2\,\sum_k\left(\frac{1}{(x^+_k)^{r-1}}-\frac{1}{(x^-_k)^{r-1}}\right).
\ee

The anomalous dimensions can be extracted from the solution of the Bethe Ansatz equations, or more efficiently, 
following the Baxter approach~\cite{Bax72}. One introduces the Baxter operator
whose eigenvalues $Q(u)$ obey a relatively simple functional equation.
If $Q(u)$ is assumed to be a polynomial, then the Baxter equation is
equivalent to the algebraic Bethe Ansatz equations for its roots
to be identified with the Bethe roots~\cite{Derkachov:1999pz,KorTrick1}.
In practice, one considers the Baxter function which is the minimal polynomial with roots equal to the Bethe roots
\be
Q(u) = \prod_{k=1}^N (u-u_k(g)).
\ee
The analysis of the multi-loop Baxter equation has been developed by A.~Belitsky and collaborators~\footnote{We are 
vary grateful to Andrei Belitsky for many helpful discussions concerning the topics of this Section.} 
in great details~\cite{MultiLoopBaxter}. In particular, analytic results for the multi-loop solution to the $SL(2)$ Baxter equation for twist 
$L=2$, $L=3$ operators can be found in~\cite{Kotikov:2008pv}~\footnote{Note a 
missing $u$ factor in the last term in Eq.~(4.9) of ~\cite{Kotikov:2008pv}}. 
This means that the loop expanded Bethe roots 
\be
u_k(g) = \sum_{\ell\ge 0} u_k^{(\ell)}\,g^{2\,\ell},
\ee
can be packaged in polynomials $Q^{\ell}(u)$ defined by 
\be
Q(u) = \prod_{1\le k\le M} (u-u_k(g)) =  \sum_{\ell\ge 0} Q^{(\ell)}(u)\,g^{2\,\ell},
\ee
and that these polynomials are explicitly known in terms of hypergeometric
functions, their derivatives and associated sums; the loop expansion of any
charge can be written in a simple way in terms of these polynomials. 
The simplest way to present the results is to write 
(for a generic set of Bethe roots, not necessarily symmetric under $u\to -u$)
\ba
q_r &=& \sum_{\ell\ge 1} q_r^{(\ell)}\,g^{2\ell}, \\
q_r^{(\ell)} &=& a_r^{(\ell)}\left(\frac{i}{2}\right)-a_r^{(\ell)}\left(-\frac{i}{2}\right),
\ea
where $a_r^{(\ell)}(u)$ is a simple function of the $F^{(\ell')}$ appearing in 
the expansion of the logarithm of the all-loop Baxter function $Q(u)$
\be
\log Q(u) = \sum_{\ell\ge 0} F^{(\ell)}(u)\,g^{2\,\ell}.
\ee
We find explicitly (for the first three charges)
\ba
a_2^{(1)} &=& 2 i F_0'(u), \\
a_2^{(2)} &=& i \left(2 F_1'(u)+F_0^{(3)}(u)\right), \\
a_2^{(3)} &=& \frac{1}{6} i \left(12 F_2'(u)+6 F_1^{(3)}(u)+F_0^{(5)}(u)\right), 
\ea

\ba
a_4^{(1)} &=& i F_0^{(3)}(u), \\
a_4^{(2)} &=& \frac{1}{4} i \left(4 F_1^{(3)}(u)+F_0^{(5)}(u)\right), \\
a_4^{(3)} &=& \frac{1}{40} i \left(40 F_2^{(3)}(u)+10 F_1^{(5)}(u)+F_0^{(7)}(u)\right),
\ea

\ba
a_6^{(1)} &=& \frac{1}{12} i F_0^{(5)}(u), \\
a_6^{(2)} &=& \frac{1}{72} i \left(6 F_1^{(5)}(u)+F_0^{(7)}(u)\right).
\ea

\subsection{A very simple sample calculation: $q_4$ at one-loop}

Let us consider the one-loop expression of the charge $q_4$. We have 
\be
q_4^{(1)} = 2\,i\,F_0^{(3)}(i/2),
\ee
where $F_0^{(3)}(u)$ is given by
\ba
F_0^{(3)}(u) = \frac{d^3}{du^3}\log Q_0(u) = 2\,\frac{(Q_0')^3}{Q_0^3}-3\frac{Q_0'Q_0''}{Q_0^2}+\frac{Q'''_0}{Q_0},
\ea
and
\be
Q_0(u) = {}_4F_3\left(\left. \begin{array}{c}
-M, \ M+1, \ \frac{1}{2}+i\,u \\
1, \ 1
\end{array}\right| 1\right).
\ee
Using (see for instance the appendices of \cite{Kotikov:2008pv})
\ba
Q_0(i/2) &=& 1, \\
Q_0'(i/2) &=& -2\,i\,S_1, \\
Q_0''(i/2) &=& 4\,(2S_{1,1}-S_2+S_{-2}), \\
Q_0'''(i/2) &=& 24\,i\,(2 S_{1,1,1}-S_{1,2}-S_{2,1}+S_{1,-2}-S_{-2,1}), \\
\ea
and the shuffle algebra for the harmonic sums~\footnote{The harmonic sums are recursively defined by 
\begin{equation*}
S_a(N) = \sum_{n=1}^N \frac{(\mbox{sign} a)^n}{n^{|a|}},\quad S_{a, b,
  \dots}(N) = \sum_{n=1}^N \frac{(\mbox{sign} a)^n}{n^{|a|}}\,S_{b, \dots}(n).
\end{equation*}
}
 we find the simple result
\be
q_4^{(1)} = -16 \left(3 S_{-3}-S_3-6 S_{-2,1}\right).
\ee

In principle, all the results for $q_{2,4,6}$ in the next sections can be obtained this way. 
However, in practice, it is much easier to 
use the maximum transcendentality Ansatz which is 
completely equivalent.

\section{$L=2$, closed formulae for multi-loops higher charges}
\label{sec:L=2}

In this section we present the obtained formulae for the first charges
$q_{2,4,6}$; the three-loops result for $q_2$ it is already
known~\cite{Kotikov:2008pv}~\cite{Kotikov:2003fb}, we recomputed it as a consistency check, and
report the result for completeness.  
For twist $L=2$ the argument of the harmonic sums is the spin $N$:  $ S_{a,b, \dots}=S_{a,b, \dots}(N)$. \\

Starting from the three-loops result for $q_2$ we have:

\ba
q_2^{(1)} &=& 8 S_1, \\
q_2^{(2)} &=& 16 \left(S_{-3}+S_3-2 S_{1,-2}-2 S_{1,2}-2 S_{2,1}\right), \\
q_2^{(3)} &=& 64 (2 S_{-5}+2 S_5-4 S_{-4,1}-2 S_{-3,-2}-S_{-3,2}-2 S_{-2,-3}-8 S_{1,-4}-4 S_{1,4}  \nonumber\\
&& -9 S_{2,-3}-5 S_{2,3}-2 S_{3,-2}-5 S_{3,2}-4 S_{4,1}+2 S_{-2,-2,1}+2 S_{-2,1,-2}\nonumber\\
&& + 8 S_{1,-3,1}+2 S_{1,-2,-2}+2 S_{1,-2,2}+12
   S_{1,1,-3}+4 S_{1,1,3}+4 S_{1,2,-2}+4 S_{1,2,2}\nonumber\\
&& 4 S_{1,3,1}+6 S_{2,-2,1}+4 S_{2,1,-2}+4 S_{2,1,2}+4 S_{2,2,1}+4 S_{3,1,1}-8 S_{1,1,-2,1})
\ea

The three-loops formulae for the first higher charge $q_4$ read:

\ba
q_4^{(1)} &=&-16 \left(3 S_{-3}-S_3-6 S_{-2,1}\right), \\
q_4^{(2)} &=&-192 (2 S_{-5}-7 S_{-4,1}-2 S_{-3,-2}-8 S_{-3,2}-S_{-2,-3}-S_{-2,3}-4 S_{1,-4}+S_{4,1}\nonumber\\
&& + 12 S_{-3,1,1}+2 S_{-2,1,-2}+2 S_{-2,1,2}+2 S_{-2,2,1}\nonumber\\
&& +8 S_{1,-3,1}+2 S_{1,-2,-2}+6 S_{1,-2,2}-8 S_{1,-2,1,1}), \\
q_4^{(3)} &=&-768 (4 S_{-7}+2 S_7-18 S_{-6,1}-7 S_{-5,-2}-34 S_{-5,2}-24
S_{-4,-3}-32 S_{-4,3}\nonumber\\
&& -18 S_{-3,-4} -10 S_{-3,4}-3 S_{-2,-5}-3 S_{-2,5}-12 S_{1,-6}-8 S_{1,6}-14
S_{2,-5}-6 S_{2,5}\nonumber\\ 
&& -2 S_{3,-4}-2 S_{4,-3} -S_{5,-2} - 4S_{6,1}+52 S_{-5,1,1}+19 S_{-4,-2,1}+21
S_{-4,1,-2}\nonumber\\
&& +59 S_{-4,1,2}+59 S_{-4,2,1} + 22 S_{-3,-3,1}+3 S_{-3,-2,-2}+7
S_{-3,-2,2}\nonumber\\ 
&& +34 S_{-3,1,-3}+22 S_{-3,1,3}+18 S_{-3,2,-2}+28 S_{-3,2,2} + 22 S_{-3,3,1}+8 S_{-2,-4,1}\nonumber\\
&& +3 S_{-2,-3,-2}+5 S_{-2,-3,2}+5 S_{-2,-2,-3}+S_{-2,-2,3}+12 S_{-2,1,-4}\nonumber\\
&& +4 S_{-2,1,4}+13 S_{-2,2,-3}+5 S_{-2,2,3}+3 S_{-2,3,-2}+5 S_{-2,3,2}+6 S_{-2,4,1}+40 S_{1,-5,1}\nonumber\\
&& +17 S_{1,-4,-2}+55 S_{1,-4,2}+38 S_{1,-3,-3}+38
   S_{1,-3,3}+16 S_{1,-2,-4}+8 S_{1,-2,4}\nonumber\\ 
&&+20 S_{1,1,-5} + 12 S_{1,1,5}+4 S_{1,2,-4}+S_{1,4,-2}+3 S_{1,4,2}+12
S_{1,5,1}+32 S_{2,-4,1} \nonumber\\ 
&& +10 S_{2,-3,-2} + 34 S_{2,-3,2}+18 S_{2,-2,-3}+18 S_{2,-2,3}+4 S_{2,1,-4}+8
S_{2,4,1}\nonumber\\ 
&& +4S_{3,-3,1}+S_{3,-2,-2} +3 S_{3,-2,2}+3 S_{4,-2,1}+S_{4,1,-2}+S_{4,1,2}+S_{4,2,1}\nonumber\\ 
&& -76 S_{-4,1,1,1}-8 S_{-3,-2,1,1}-24 S_{-3,1,-2,1} - 28 S_{-3,1,1,-2}-24 S_{-3,1,1,2}\nonumber\\ 
&& -24 S_{-3,1,2,1}-24 S_{-3,2,1,1}-8 S_{-2,-3,1,1}-6 S_{-2,-2,-2,1} - 4 S_{-2,-2,1,-2}\nonumber\\ 
&&-4 S_{-2,-2,1,2}-4 S_{-2,-2,2,1}-16 S_{-2,1,-3,1}-2 S_{-2,1,-2,-2}-6 S_{-2,1,-2,2}\nonumber\\
&& -20 S_{-2,1,1,-3}-4 S_{-2,1,1,3}-8 S_{-2,1,2,-2}-4 S_{-2,1,2,2}-4 S_{-2,1,3,1}-10 S_{-2,2,-2,1}\nonumber\\
&& -8
   S_{-2,2,1,-2}-4 S_{-2,2,1,2}-4 S_{-2,2,2,1}-4 S_{-2,3,1,1}-76 S_{1,-4,1,1}-32 S_{1,-3,-2,1}\nonumber\\
&& -28 S_{1,-3,1,-2}-60 S_{1,-3,1,2}-60 S_{1,-3,2,1}-20 S_{1,-2,-3,1}-4 S_{1,-2,-2,-2}-8 S_{1,-2,-2,2}\nonumber\\
&& -24 S_{1,-2,1,-3}-16
   S_{1,-2,1,3}-12 S_{1,-2,2,-2}-20 S_{1,-2,2,2}-16 S_{1,-2,3,1}-40 S_{1,1,-4,1}\nonumber\\
&& -16 S_{1,1,-3,-2}-40 S_{1,1,-3,2}-32 S_{1,1,-2,-3}-24 S_{1,1,-2,3}-16 S_{1,1,4,1}-8 S_{1,2,-3,1}\nonumber\\
&& -2 S_{1,2,-2,-2}-6 S_{1,2,-2,2}-4
   S_{1,4,1,1}-48 S_{2,-3,1,1}-18 S_{2,-2,-2,1}-10 S_{2,-2,1,-2} \nonumber\\ 
&& -30 S_{2,-2,1,2} -30 S_{2,-2,2,1}-8 S_{2,1,-3,1}-2 S_{2,1,-2,-2}-6
S_{2,1,-2,2}-4 S_{3,-2,1,1}\nonumber\\ 
&&-4 S_{4,1,1,1}+8 S_{-2,-2,1,1,1} + 8 S_{-2,1,-2,1,1}+16
   S_{-2,1,1,-2,1}+8 S_{-2,1,1,1,-2}\nonumber\\
&& +64 S_{1,-3,1,1,1}+8 S_{1,-2,-2,1,1}+16 S_{1,-2,1,-2,1} + 16 S_{1,-2,1,1,-2}+16 S_{1,-2,1,1,2}\nonumber\\
&& + 16 S_{1,-2,1,2,1}+16 S_{1,-2,2,1,1}+48 S_{1,1,-3,1,1}+32 S_{1,1,-2,-2,1}\nonumber\\
&& +16
   S_{1,1,-2,1,-2}+32 S_{1,1,-2,1,2}+32 S_{1,1,-2,2,1}+8
   S_{1,2,-2,1,1}\nonumber\\ 
&& +40 S_{2,-2,1,1,1}+8 S_{2,1,-2,1,1} - 32 S_{1,1,-2,1,1,1})
\ea


It is well known that the all-loop Bethe equations must be "completed" by 
a dressing factor~\cite{Beisert:2006ez}~\cite{Beisert:2005wv}, that starts to
contribute, at weak coupling, from the four-loops 
term: then, at four-loops we have that the $q_4$ is a sum of two parts:
\be
q_4 = \cdots + g^8\,(q_4^{(4, \rm rational)} + \zeta_3\,q_4^{(4, \rm dressing)}) + \cdots .
\ee
We report here only the $\zeta_3$ part, which comes from the dressing phase
and is a combination of transcendentality 6 harmonic sums. We have found

\ba
q_4^{(4, \rm dressing)} &=& -3072 (S_{-6}-S_6-S_{-5,1}+2 S_{-4,2}+2 S_{-3,-3}+2 S_{-3,3}+2 S_{-2,-4}
\sxx -S_{1,-5}+S_{1,5}+2 S_{4,2}+3 S_{5,1}-4 S_{-4,1,1}-4 S_{-3,-2,1}-10 S_{-3,1,2}-10 S_{-3,2,1}
\sxx -6 S_{-2,-3,1}-4 S_{-2,-2,2}-2 S_{-2,1,-3}-4 S_{-2,1,3}-8 S_{-2,2,2}-4S_{-2,3,1}
\sxx -2 S_{1,-3,2} - 2 S_{1,-2,-3}-2 S_{1,4,1}-4 S_{4,1,1}+20 S_{-3,1,1,1}
\sxx +8 S_{-2,-2,1,1}+4 S_{-2,1,-2,1} + 16 S_{-2,1,1,2}+16 S_{-2,1,2,1}
\sxx +16 S_{-2,2,1,1}+4 S_{1,-3,1,1}+4 S_{1,-2,-2,1} +4 S_{1,-2,1,2} + 4S_{1,-2,2,1}
\sxx -32 S_{-2,1,1,1,1}-8 S_{1,-2,1,1,1}).
\ea

For the $q_6$ charge we show the two-loops result:

\ba
q_6^{(1)} &=& 64 (2 S_5-5 S_{-4,1}-10 S_{-3,2}-5 S_{-2,-3}-5 S_{-2,3}-5 S_{4,1}+20 S_{-3,1,1}+10 S_{-2,-2,1}\nonumber\\
&& +20 S_{-2,1,2}+20 S_{-2,2,1}-40 S_{-2,1,1,1}), \\
q_6^{(2)} &=& 640 (4 S_7-4 S_{-6,1}-2 S_{-5,-2}-18 S_{-5,2}-13 S_{-4,-3}-25 S_{-4,3}-28 S_{-3,-4}-16 S_{-3,4}
\sxx -8 S_{-2,-5}-4 S_{-2,5}-8 S_{1,6}-S_{4,-3}-S_{4,3}-2 S_{5,-2}-18 S_{5,2}-16 S_{6,1}+32 S_{-5,1,1}
\sxx +16 S_{-4,-2,1}+10 S_{-4,1,-2}+62 S_{-4,1,2}+62 S_{-4,2,1}+60 S_{-3,-3,1}+4 S_{-3,-2,-2}
\sxx +36 S_{-3,-2,2}+36 S_{-3,1,-3}+52 S_{-3,1,3}+16 S_{-3,2,-2}+72 S_{-3,2,2}+52 S_{-3,3,1}
\sxx +22 S_{-2,-4,1}+4 S_{-2,-3,-2}+20 S_{-2,-3,2}+2 S_{-2,-2,-3}+2 S_{-2,-2,3}+24 S_{-2,1,-4}
\sxx +16 S_{-2,1,4}+20 S_{-2,2,-3}+28 S_{-2,2,3}+8 S_{-2,3,-2}+28 S_{-2,3,2}+14 S_{-2,4,1}
\sxx +8 S_{1,-5,1}+2 S_{1,-4,-2}+18 S_{1,-4,2}+16 S_{1,-3,-3}+16 S_{1,-3,3}+24 S_{1,-2,-4}
\sxx +8 S_{1,-2,4}+2 S_{1,4,-2}+18 S_{1,4,2}+24 S_{1,5,1}+2 S_{4,1,-2}+2 S_{4,1,2}+2 S_{4,2,1}
\sxx +32 S_{5,1,1}-104 S_{-4,1,1,1}-64 S_{-3,-2,1,1}-40 S_{-3,1,-2,1}-32 S_{-3,1,1,-2}-112 S_{-3,1,1,2}
\sxx -112 S_{-3,1,2,1}-112 S_{-3,2,1,1}-32 S_{-2,-3,1,1}-4 S_{-2,-2,1,-2}-4 S_{-2,-2,1,2}-4 S_{-2,-2,2,1}
\sxx -40 S_{-2,1,-3,1}-4 S_{-2,1,-2,-2}-20 S_{-2,1,-2,2}-40 S_{-2,1,1,-3}-40 S_{-2,1,1,3}
\sxx -24 S_{-2,1,2,-2} - 48 S_{-2,1,2,2}-40 S_{-2,1,3,1}-16 S_{-2,2,-2,1}-24 S_{-2,2,1,-2}-48 S_{-2,2,1,2}
\sxx -48 S_{-2,2,2,1}-40 S_{-2,3,1,1}-32 S_{1,-4,1,1}-24 S_{1,-3,-2,1}-8 S_{1,-3,1,-2}-40 S_{1,-3,1,2}
\sxx -40 S_{1,-3,2,1}-56 S_{1,-2,-3,1}-4 S_{1,-2,-2,-2}-36 S_{1,-2,-2,2}-24 S_{1,-2,1,-3}-24 S_{1,-2,1,3}
\sxx -8 S_{1,-2,2,-2}-32 S_{1,-2,2,2}-24 S_{1,-2,3,1}-32 S_{1,4,1,1}+160 S_{-3,1,1,1,1}+32 S_{-2,1,-2,1,1}
\sxx +32 S_{-2,1,1,-2,1}+48 S_{-2,1,1,1,-2}+48 S_{-2,1,1,1,2}+48 S_{-2,1,1,2,1}+48 S_{-2,1,2,1,1}
\sxx +48 S_{-2,2,1,1,1}+64 S_{1,-3,1,1,1}+64 S_{1,-2,-2,1,1}+32 S_{1,-2,1,-2,1}+16 S_{1,-2,1,1,-2}
\sxx +48 S_{1,-2,1,1,2}+48 S_{1,-2,1,2,1}+48 S_{1,-2,2,1,1}-64 S_{1,-2,1,1,1,1})
\ea

\section{$L=3$, closed formulae for multi-loops higher charges}
\label{sec:L=3}

As in the previous section we report also the known result for $q_2$~\cite{Beccaria:2007cn}~\cite{Kotikov:2007cy}.
For $L=3$ the argument of the harmonic sums is half the spin $N/2$.
Starting from $q_2$, we have:

\ba
q_2^{(1)} &=& 8 S_1, \\
q_2^{(2)} &=& 8 \left(S_3-2 S_{1,2}-2 S_{2,1}\right), \\
q_2^{(3)} &=& 8 (S_5-2 S_{1,4}-6 S_{2,3}-10 S_{3,2}-6 S_{4,1}+8 S_{1,2,2}+8 S_{1,3,1}
\sxx +8 S_{2,1,2}+8 S_{2,2,1}+8 S_{3,1,1})
\ea

For twist-3 the three-loops result for $q_4$ is

\ba
q_4^{(1)} &=& 16 S_3, \\
q_4^{(2)} &=& 96 \left(S_5-2 S_{1,4}-2 S_{4,1}+2 S_{1,3,1}\right), \\
q_4^{(3)} &=& 
     48 (9 S_7-32 S_{1,6}-36 S_{2,5}-4 S_{3,4}-4 S_{4,3}-36 S_{5,2}-32 S_{6,1}+56 S_{1,1,5}
\sxx +8 S_{1,2,4}+4 S_{1,3,3}+56 S_{1,4,2}+76 S_{1,5,1}+8 S_{2,1,4}+32 S_{2,3,2}+56 S_{2,4,1}
\sxx +4 S_{3,3,1}+8 S_{4,1,2}+8 S_{4,2,1}+56 S_{5,1,1}-48 S_{1,1,3,2}-80 S_{1,1,4,1}
\sxx -8 S_{1,2,3,1}-8 S_{1,3,1,2}-8 S_{1,3,2,1}-80 S_{1,4,1,1}-8 S_{2,1,3,1}
\sxx -48 S_{2,3,1,1}+64 S_{1,1,3,1,1})
\ea

and the two-loops for $q_6$

\ba
q_6^{(1)} &=& -32 \left(S_5-5 S_{3,2}-5 S_{4,1}+10 S_{3,1,1}\right), \\
q_6^{(2)} &=&  
     48 (9 S_7-32 S_{1,6}-36 S_{2,5}-4 S_{3,4}-4 S_{4,3}-36 S_{5,2}-32 S_{6,1}+56 S_{1,1,5}+8 S_{1,2,4}
\sxx +4 S_{1,3,3}+56 S_{1,4,2}+76 S_{1,5,1}+8 S_{2,1,4}+32 S_{2,3,2}+56 S_{2,4,1}+4 S_{3,3,1}+8 S_{4,1,2}
\sxx +8 S_{4,2,1}+56 S_{5,1,1}-48 S_{1,1,3,2}-80 S_{1,1,4,1}-8 S_{1,2,3,1}-8 S_{1,3,1,2}-8 S_{1,3,2,1}
\sxx -80 S_{1,4,1,1}-8 S_{2,1,3,1}-48 S_{2,3,1,1}+64 S_{1,1,3,1,1})
\ea

\section{Large spin expansions}
\label{sec:largespin}

The large spin limit of the anomalous dimensions has been intensively 
investigated in the recent past, due to its relevance in the comparison with
the string theory results.\\

In this section we report the large spin expansion for 
the previously obtained formulae: it has the usual form 
$q_{r, L}(N) = \log(N) f_r(g) + B_{r, L}(g) + ...$ where the coefficient of the 
$\log (N)$ is $L$-independent whereas the constant, (the \emph{virtual scaling function}),
is twist dependent. The functions $f_r(g)$ can be computed at all orders from the solution of the
BES equation~\cite{Beisert:2006ez}, while the derivation of $B_{r,L}$
requires an integral equation which is valid at order $\mathcal{O}(N^0)$~\cite{Bombardelli:2008ah,Fioravanti:2009xt,Fioravanti:2009xn,Freyhult:2009my}. 
This expansion allows a non trivial check of our results and a comparison
with the very useful results of~\cite{Bombardelli:2008ah} (section \ref{sec:checks}).\\ 

Again, we always report the three-loops $q_2$ for better comparisons. $M$ is
defined as $M=N$ for twist $L=2$ and $M=N/2$ for $L=3$; To simplify the
notation we omit a $e^{\gamma_E}$ factor in the 
argument of all logarithms; this reabsorbs all Euler-Gamma constants, as
usual.\\

For twist $L=2$ we obtain for $q_2,q_4$ and $q_6$ respectively:\\

{\bf $L=2$, $q_2$}

\ba
q_2^{(1)} &=& 8\,\log M + \frac{4}{M}-\frac{2}{3\,M^2} + \cdots, \\
q_2^{(2)} &=& -\frac{8}{3}\,\pi^2\,\log M -24\,\zeta_3 + \left(32\,\log M-\frac{4\pi^2}{3}\right)\,\frac{1}{M} 
\sxx + \left(-16\,\log M+24+\frac{2\pi^2}{9}\right)\,\frac{1}{M^2} + \cdots, \\
q_2^{(3)} &=& \frac{88}{45}\, \pi ^4\, \log M+160\, \zeta_5+\frac{16}{3}\, \pi ^2 \,\zeta_3+\left(-\frac{64}{3} \pi ^2 \log M-96\, \zeta_3+\frac{44 \pi ^4}{45}\right)\,\frac{1}{M}
\sxx 
+\left(-64 \log^2 M+(128 + \frac{16}{3} \pi ^2) \log M+48 \zeta_3-\frac{22 \pi ^4}{135}-\frac{32 \pi ^2}{3}\right) \,\frac{1}{M^2} + \cdots
\ea

{\bf $L=2$, $q_4$}

\ba
q_4^{(1)} &=& -8 \zeta_3+(48 \log M-8) \frac{1}{M^2} + \cdots, \\
q_4^{(2)} &=& \frac{8}{15} \pi ^4 \log M+120 \zeta_5+\frac{4 \pi ^4}{15 M}
\sxx +\left(-16 \pi ^2 \log M-144 \zeta_3-\frac{2 \pi ^4}{45}\right)\frac{1}{M^2} + \cdots, \\
q_4^{(3)} &=& -\frac{296}{315} \pi ^6 \log M-1512 \zeta_7+16 \pi ^2 \zeta_5-\frac{8}{5} \pi ^4 \zeta_3+\left(\frac{32}{15} \pi ^4 \log M-\frac{148 \pi ^6}{315}\right)\frac{1}{M}
\sxx +\left(\frac{128}{3} \pi ^2
   \log ^3 M+\frac{32}{3} \pi ^4 \log M+960 \zeta_5+\right.
\sxx \left.\qquad\qquad\qquad\qquad\qquad\qquad + \frac{112}{3} \pi ^2 \zeta_3+\frac{74 \pi ^6}{945}+\frac{8 \pi ^4}{5}\right)\frac{1}{M^2} + \cdots, \\
q_4^{(4, \rm dressing)} &=& 768 \log M \zeta_5+\frac{384 \zeta_5}{M}+\left(2048 \log ^4 M-512 \zeta_3 \log M-64 \zeta_5\right) \frac{1}{M^2}
+\cdots.
\ea

{\bf $L=2$, $q_6$}

\ba
q_6^{(1)} &=& 8 \zeta_5+\left(-\frac{640}{3} \log ^3 M-\frac{80 \zeta_3}{3}\right) \frac{1}{M^2} + \cdots, \\
q_6^{(2)} &=& -\frac{16}{189} \pi ^6 \log M-280 \zeta _7-\frac{8 \pi ^6}{189 M}
\sxx +\left(\frac{640}{3} \pi ^2 \log ^3 M+1920 \zeta _3 \log ^2 M+\frac{16}{9} \pi ^4 \log M+400
   \zeta _5+\frac{4 \pi ^6}{567}\right) \frac{1}{M^2} + \cdots
\ea

And for twist 3:
{\bf $L=3$, $q_2$}

\ba
q_2^{(1)} &=& 8 \log M+\frac{4}{M}-\frac{2}{3 M^2} + \cdots, \\
q_2^{(2)} &=& -\frac{8}{3} \pi ^2 \log M-8 \zeta _3+\left(16 \log M-\frac{4 \pi ^2}{3}\right)\frac{1}{M}
\sxx +\left(-8 \log M+\frac{2 \pi ^2}{9}+12\right) \frac{1}{M^2} + \cdots, \\
q_2^{(3)} &=& \frac{88}{45} \pi ^4 \log M+\frac{8 \pi ^2 \zeta _3}{3}-8 \zeta _5+\left(-\frac{32}{3} \pi ^2 \log M-16 \zeta _3+\frac{44 \pi ^4}{45}\right)\frac{1}{M}
\sxx \! +\left(-16 \log ^2 M+(\frac{16}{3}
   \pi ^2 +32) \log M+8 \zeta _3-\frac{22 \pi ^4}{135}-\frac{20 \pi ^2}{3}+8\right) \frac{1}{M^2} + \cdots
\ea

{\bf $L=3$, $q_4$}

\ba
q_4^{(1)} &=& 16 \zeta _3-\frac{8}{M^2} + \cdots, \\
q_4^{(2)} &=& \frac{8}{15} \pi ^4 \log M-192 \zeta _5+\frac{4 \pi ^4}{15 M}+\left(48 M-\frac{2 \pi ^4}{45}+48\right) \frac{1}{M^2} + \cdots, \\
q_4^{(3)} &=& -\frac{296}{315} \pi ^6 \log M-\frac{8 \pi ^4 \zeta _3}{15}+2304 \zeta _7+\left(\frac{16}{15} \pi ^4 \log M-\frac{148 \pi ^6}{315}\right)\frac{1}{M} \\
&&\!\!\!\! +\left(-192 \log ^2 M-(\frac{8}{15} \pi^4 +16 \pi ^2 +576) \log M -48 \zeta _3+\frac{74 \pi ^6}{945}+\frac{4 \pi ^4}{5}-576\right) \frac{1}{M^2} + \cdots\nonumber
\ea

{\bf $L=3$, $q_6$}

\ba
q_6^{(1)} &=& -112 \zeta _5+\left(80 \log ^2 M+80 \log M+40\right) \frac{1}{M^2} +\cdots, \\
q_6^{(2)} &=& 2120 \zeta _7-\frac{16}{189} \pi ^6 \log M-\frac{8 \pi ^6}{189 M}
\sxx +\left(-\frac{640}{3} \log ^3 M+\left(-640-\frac{160 \pi ^2}{3}\right) \log ^2 M+\right. 
\sxx \left. \left(-160 \zeta
   _3-\frac{80 \pi ^2}{3}-960\right) \log M-\frac{80 \zeta _3}{3}+\frac{4 \pi ^6}{567}-640\right) \frac{1}{M^2} + \cdots
\ea

\subsection{The NLO terms at large spin from integral equations}
\label{sec:checks}

From the exact results of the previous section  we can immediately extract the
NLO large spin expansion of the charges at twist $L$
\be
q_{r, L}(N, g) = f_r(g)\,\log N + B_{r, L}(g) + o(N), 
\ee
where $f_r(g)$ is a universal function (it does not depend on the twist $L$) and takes the values 
\ba
\label{eq:scaling}
f_4(g) &=& 0\cdot g^2 + \frac{8\,\pi^4}{15}\,g^4-\frac{296\,\pi^6}{315}\,g^6 + \cdots, \\
f_6(g) &=& 0\cdot g^2 -\frac{16\,\pi^6}{189}\,g^4 + \cdots, \nonumber
\ea
and the so celled {\em virtual scaling function} $B_{r, L}(g)$ takes the values 
\ba
\label{eq:vir1}
B_{4, 2}(g) &=& -8\,\zeta_3\,g^2+120\,\zeta_5\,g^4+\left(-1512\,\zeta_7+16\,\pi^2\,\zeta_5-\frac{8}{5}\,\pi^4\,\zeta_3\right)\,g^6 + \cdots, \\
B_{4, 3}(g) &=& 16\,\zeta_3\,g^2-192\,\zeta_5\,g^4+\left(2304\,\zeta_7-\frac{8}{15}\,\pi^4\,\zeta_3\right)\,g^6 + \cdots, \nonumber
\ea
and
\ba
\label{eq:vir2}
B_{6, 2}(g) &=& 8\,\zeta_5\,g^2-280\,\zeta_7\,g^6 + \cdots, \\ 
B_{6, 3}(g) &=& -112\,\zeta_5\,g^2+2120\,\zeta_7\,g^6 + \cdots. \nonumber
\ea
It is worthwhile to note that our results are valid for any $N$ and, 
for instance, they provide the full large $N$ expansion of the charges, not 
just the above NLO coefficients.

\subsection{Check of the scaling functions $f_{r}(g)$}

As a first check of our expressions, we found that \refeq{scaling} are in
perfect agreement with an alternative (and more efficient) calculation of $f_r(g)$ from the BES
equation:  indeed, it is enough to plug the perturbative BES 
density $\widehat\sigma(t)$ defined in~\cite{Beisert:2006ez}
in the following continuum limit expression of the charges
\be
\label{eq:bes}
f_r(g) = -64\,g^4\,\frac{r-1}{(i\,g)^{r-2}}\int_0^\infty dt\,\widehat{\sigma}(t)\,\frac{J_{r-1}(2\,g\,t)}{2\,g\,t}
\ee
Note that this expression gives the contributions beyond one-loop which is
enough for the higher charges, but which must be completed with the
one-loop contribution in the case of the energy $q_2$.\\

For instance, we find from \refeq{bes}

\ba
f_2(g) &=& -\frac{8 \pi ^2 g^4}{3}+\frac{88 \pi ^4 g^6}{45}+\left(-64 \zeta _3^2-\frac{584 \pi ^6}{315}\right) g^8
\sxx +\left(\frac{128}{3} \pi ^2 \zeta _3^2+1280 \zeta _5 \zeta _3+\frac{28384 \pi ^8}{14175}\right)
   g^{10}+O\left(g^{12}\right), \\
f_4(g) &=& \frac{8 \pi ^4 g^4}{15}-\frac{296 \pi ^6 g^6}{315}+\left(384 \zeta _3 \zeta _5+\frac{1304 \pi ^8}{945}\right) g^8 \\
&& +\left(-\frac{64}{15} \pi ^4 \zeta _3^2-128 \pi ^2 \zeta _5 \zeta _3-6720 \zeta _7 \zeta _3-3840 \zeta
   _5^2-\frac{303416 \pi ^{10}}{155925}\right) g^{10}+O\left(g^{12}\right)\nonumber, \\
f_6(g) &=& -\frac{16 \pi ^6 g^4}{189}+\frac{752 \pi ^8 g^6}{2835}+\left(-960 \zeta _3 \zeta _7-\frac{7504 \pi ^{10}}{13365}\right) g^8\\
&&\!\! +\left(\frac{128}{189} \pi ^6 \zeta _3^2+320 \pi ^2 \zeta _7 \zeta _3+26880 \zeta _9 \zeta
   _3+9600 \zeta _5 \zeta _7+\frac{43466152 \pi ^{12}}{42567525}\right) g^{10}+O\left(g^{12}\right), \nonumber \\
f_8(g) &=& \frac{8 \pi ^8 g^4}{675}-\frac{1288 \pi ^{10} g^6}{22275}+\left(1792 \zeta _3 \zeta _9+\frac{15280024 \pi ^{12}}{91216125}\right) g^8\\
&& +\left(-\frac{64}{675} \pi ^8 \zeta _3^2-\frac{1792}{3} \pi ^2 \zeta _9 \zeta _3-73920
   \zeta _{11} \zeta _3-  \nonumber\right.\\
&& \left. \qquad\qquad\qquad\qquad\qquad\qquad - 17920 \zeta _5 \zeta _9-\frac{105442408 \pi ^{14}}{273648375}\right) g^{10}+O\left(g^{12}\right)\nonumber
\ea

and so on.
Notice that we have computed the four-loops ($\mathcal{O}(g^8)$) dressing constant in $f_4$
which is $+768\zeta_3\zeta_5$. Comparing with the above expansion
one can see that the rational part must be $-384\zeta_3\zeta_5$ and 
the dressing contribution just flips the sign of the naive term, precisely as for 
the scaling function $f_2(g)$ itself, where the term $-64 \zeta _3^2 g^8 $ comes from
$(64^{\textrm{(rational)}}-128^{\textrm{(dressing)}})\zeta_3^2 g^8$ \cite{Beisert:2006ez}.

\subsection{Check of the virtual scaling functions $B_{r, L}(g)$}

As a further test \refeq{vir1} and \refeq{vir2} can be checked against the 
three-loops calculation of the virtual scaling functions reported 
in~\cite{Bombardelli:2008ah}. Their formulae, written consistently with 
our notation, provide the following result
\ba
B_{r, L}^{(1)} &=& 2 i^r \left(2^r L-4 L-2^{r+1}+4\right) \zeta _{r-1}, \\
B_{r, L}^{(2)} &=& 4 i^{3 r} (r-1) \left(-2^{r+1} L+r L+2 L+2^{r+2}-2\right) \zeta _{r+1}, \\
B_{r, L}^{(3)} &=& 16 i^r (2 L-7) (r-1) \zeta _3 \zeta _r  -\frac{4}{3} i^r (L-3) \pi ^2 (r-1) r \zeta _{r+1}
\sxx i^r (r-1) (r+2) \left(-L r^2-5 L r-2^{r+5}+2^{r+4} L-8 L+4\right) \zeta _{r+3}
\ea
Replacing $r=4, 6$ and $L=2, 3$ we find perfect agreement with Eqs.~(\ref{eq:vir1},\ref{eq:vir2}).

\section{Reciprocity properties of the higher charges}
\label{sec:reciprocity}

Following and generalizing the treatment in~\cite{Beccaria:2009vt}, 
for a $L=2$ charge $q_r(N)$ the reciprocity condition can be defined as constraint on the
large spin expansion: introducing the function $P_r(N)$ as 
\be
q_r(N) = P_r(N + \frac{1}{2} q_2(N)),
\label{eq:P}
\ee
reciprocity implies that  the large $N$ expansion of $P_r$ involves  integer
inverse powers of $N\,(N+1)$ only.
A similar definition holds for twist-3 with $N/2$ in place of $N$.\\

The definition of the function $P_r$ for the higher charges is not "trivial":
at a first sight one could try to generalize the prescription for $q_2$ simply 
replacing the argument with $N + \frac{1}{2} q_r(N)$, but this choice is not 
reciprocity respecting.
In our definition of  $P_r$ the argument is 
$N + \frac{1}{2} q_2(N)$. Additional reasons for this choice have already been  discussed in the Introduction. 

In the following subsections we list 
the perturbative functions $P_r$ for the various cases: they are combinations of harmonic sums with 
various transcendentality. It is convenient to rewrite the results in terms 
of the $\Omega$ basis discussed in~\cite{Beccaria:2009vt}, where the check of
the reciprocity is straightforward:
{\em reciprocity holds iff the occurring $\Omega$ have odd positive or even negative indices}.\\

{\bf $L=2$, three-loops reciprocity of $q_4$}

\ba
P_4^{(1)} &=&16 \left(\Omega _3+6 \Omega _{-2,1}\right), \\
P_4^{(2)} &=& -\frac{16}{5} (\pi ^4 \Omega _1+120 \Omega _{-4,1}+20 \pi ^2 \Omega _{-2,1}+60 \Omega _{-2,3}+60 \Omega _{1,-4}+20 \pi ^2 \Omega _{1,-2}
\sxx+120 \Omega _{-2,1,-2} +120 \Omega _{1,-2,-2}-480 \Omega _{1,-2,1,1}), \\
P_4^{(3)} &=& \frac{32}{15} (180 \zeta (3) \Omega _{-4}+2 \pi ^6 \Omega _1+3 \pi ^4 \Omega _3-30 \pi ^2 \Omega _5-720 \Omega _7+900 \Omega _{-6,1}+240 \pi ^2 \Omega _{-4,1}
\sxx +540 \Omega _{-4,3}+30 \pi ^4 \Omega _{-2,1}+60 \pi ^2 \Omega _{-2,3}+720\Omega _{1,-6}+240 \pi ^2 \Omega _{1,-4}+36 \pi ^4 \Omega _{1,-2}
\sxx +180 \Omega _{3,-4} +60 \pi ^2 \Omega _{3,-2}-180 \Omega _{5,-2}+2520\Omega _{-4,-2,1}+2160 \Omega _{-4,1,-2}
\sxx +1080 \Omega _{-2,-4,1}+360 \Omega _{-2,-2,3} + 1800 \Omega _{-2,1,-4} + 120 \pi ^2 \Omega _{-2,1,-2}
\sxx +1080 \Omega _{-2,3,-2}+1440 \Omega _{1,-4,-2}+2160 \Omega _{1,-2,-4}
\sxx +240 \pi ^2 \Omega _{1,-2,-2}+1440 \Omega _{1,1,5}+2160 \Omega_{1,5,1}+360 \Omega _{3,-2,-2}+720 \Omega _{5,1,1}
\sxx -1440 \Omega _{-4,1,1,1}+2160 \Omega _{-2,-2,-2,1}
\sxx +1440 \Omega _{-2,-2,1,-2}+720 \Omega _{-2,1,-2,-2}-2880 \Omega _{1,-4,1,1}
\sxx +1440 \Omega _{1,-2,-2,-2}-960 \pi ^2 \Omega _{1,-2,1,1}-1440 \Omega _{1,-2,1,3}-1440 \Omega _{1,-2,3,1}-1440 \Omega _{1,1,-4,1}
\sxx -960 \pi ^2 \Omega _{1,1,-2,1}-1440 \Omega _{1,1,-2,3}-1440 \Omega _{3,-2,1,1}-2880 \Omega _{-2,-2,1,1,1}
\sxx -2880 \Omega _{-2,1,-2,1,1}-5760 \Omega _{-2,1,1,-2,1}-2880 \Omega_{-2,1,1,1,-2}-2880 \Omega _{1,-2,-2,1,1}
\sxx -5760 \Omega _{1,-2,1,-2,1} -5760 \Omega _{1,-2,1,1,-2}-11520 \Omega_{1,1,-2,-2,1}
\sxx -5760 \Omega _{1,1,-2,1,-2}+11520 \Omega _{1,1,-2,1,1,1} +360 \Omega_{1,1} \zeta (5)-240 \pi ^2 \Omega _{1,1} \zeta (3)
\sxx-720 \Omega _{-2,1,1} \zeta (3)-720 \Omega _{1,-2,1} \zeta (3) -720 \Omega _{1,1,-2} \zeta (3))
\ea

{\bf $L=2$, four-loops reciprocity of the dressing part of $q_4$}

\ba
P_4^{(4, \rm dressing)} &=& 3072 \Omega _{-6}+3072 \Omega _{-2,-4}+3072 \Omega _{5,1}-18432 \Omega _{-4,1,1}
\sxx -12288 \Omega _{-2,1,3}-12288 \Omega _{-2,3,1}-6144 \Omega _{1,-4,1}-6144 \Omega _{1,-2,3}
\sxx -24576 \Omega _{-2,-2,1,1}-12288 \Omega _{-2,1,-2,1}-12288 \Omega _{1,-2,-2,1}
\sxx +98304 \Omega _{-2,1,1,1,1}+24576 \Omega _{1,-2,1,1,1}.
\ea

{\bf $L=2$, two-loops reciprocity of $q_6$}

\ba
P_6^{(1)} &=& -32 \left(\Omega _5-10 \Omega _{-4,1}-10 \Omega _{-2,3}-20 \Omega _{-2,-2,1}+80 \Omega _{-2,1,1,1}\right), \\
P_6^{(2)} &=& -\frac{32}{63} (\pi ^6 \Omega _1-210 \pi ^2 \Omega _5-3150 \Omega _7+5040 \Omega _{-6,1}+1260 \pi ^2 \Omega _{-4,1}
\sxx +3780 \Omega _{-4,3}+42 \pi ^4 \Omega _{-2,1}+840 \pi ^2 \Omega _{-2,3}+2520 \Omega _{-2,5}+1260\Omega _{1,-6}
\sxx +420 \pi ^2 \Omega _{1,-4}+42 \pi ^4 \Omega _{1,-2}-1260 \Omega _{5,-2}+10080 \Omega _{-4,-2,1}+7560 \Omega _{-4,1,-2}
\sxx +10080 \Omega _{-2,-4,1}+840 \pi ^2 \Omega _{-2,-2,1}+2520 \Omega _{-2,-2,3}+12600 \Omega _{-2,1,-4}
\sxx +840 \pi ^2 \Omega _{-2,1,-2}+5040 \Omega _{-2,3,-2}+2520 \Omega _{1,-4,-2}+12600 \Omega _{1,-2,-4}
\sxx +840 \pi ^2 \Omega _{1,-2,-2}+10080 \Omega _{1,5,1}+10080 \Omega _{5,1,1}-20160 \Omega _{-4,1,1,1}
\sxx +5040 \Omega _{-2,-2,1,-2}+5040 \Omega _{-2,1,-2,-2}-10080 \pi ^2 \Omega _{-2,1,1,1}-10080 \Omega _{-2,1,1,3}
\sxx -10080 \Omega _{-2,1,3,1}-10080 \Omega _{-2,3,1,1}-10080 \Omega _{1,-4,1,1}+5040 \Omega_{1,-2,-2,-2}
\sxx -3360 \pi ^2 \Omega _{1,-2,1,1}-10080 \Omega _{1,-2,1,3}-10080 \Omega _{1,-2,3,1}-40320 \Omega _{-2,1,-2,1,1}
\sxx -40320 \Omega _{-2,1,1,-2,1}-60480 \Omega _{-2,1,1,1,-2}-80640 \Omega _{1,-2,-2,1,1}
\sxx -40320 \Omega _{1,-2,1,-2,1}-20160 \Omega _{1,-2,1,1,-2}+80640 \Omega _{1,-2,1,1,1,1})
\ea

{\bf $L=3$, three-loops reciprocity of $q_4$}

\ba
P_4^{(1)} &=& 16 \Omega _3, \\
P_4^{(2)} &=& -\frac{16}{15} \left(\pi ^4 \Omega _1+45 \Omega _5-180 \Omega _{1,3,1}\right), \\
P_4^{(3)} &=&\frac{16}{315} (16 \pi ^6 \Omega _1+21 \pi ^4 \Omega _3+315 \pi ^2 \Omega _5+3780 \Omega _7
\sxx -2520 \pi ^2 \Omega _{1,1,3}-7560 \Omega _{1,1,5}-1260 \pi ^2 \Omega _{1,3,1}-3780 \Omega _{1,3,3}
\sxx -11340 \Omega_{1,5,1}-3780 \Omega _{3,3,1}-7560 \Omega _{5,1,1}+60480 \Omega _{1,1,3,1,1}
\sxx -22680 \Omega _{1,1} \zeta (5)+2520 \pi ^2 \Omega _{1,1} \zeta (3))
\ea

{\bf $L=3$, two-loops reciprocity of $q_6$}

\ba
P_6^{(1)} &=& 16 \left(3 \Omega _5-20 \Omega _{3,1,1}\right), \\
P_6^{(2)} &=& -\frac{8}{63} (\pi ^6 \Omega _1+210 \pi ^2 \Omega _5+1890 \Omega _7-840 \pi ^2 \Omega _{1,3,1}-5040 \Omega _{1,3,3}
\sxx -5040 \Omega _{1,5,1}-1680 \pi ^2 \Omega _{3,1,1}-2520 \Omega _{3,1,3}-2520 \Omega _{3,3,1}
\sxx -7560   \Omega _{5,1,1}+40320 \Omega _{1,3,1,1,1})
\ea

As one can see from the previous formulae we find a very strong indication that reciprocity extends to the higher charges.








\section{Strong coupling}
\label{sec:strong}

\subsection{Reciprocity of the energy}

We consider the semiclassical $\mathfrak{sl}(2)$ folded string, the dual partner of 
the single trace operators previously considered in the gauge side.
As discussed in the Introduction this solution describes a string rotating in
$\ads$ and stretched in the radial direction of $AdS_5$~\cite{Frolov:2002av};
the energy ${\cal E} = E/\sqrt{\lambda}$, spin ${\cal S} = S/\sqrt\lambda$, and angular momentum $\J = J/\sqrt\lambda$
are parametrically related by~\cite{Beccaria:2008tg}
\ba
\sqrt{\kappa^2-\J^2} &=& \frac{1}{\sqrt\eta}\,{}_2F_1\left(\frac{1}{2}, \frac{1}{2}, 1, -\frac{1}{\eta}\right), \\
\omega^2-\J^2 &=& (1+\eta)\,(\kappa^2-\J^2), 
\ea
and 
\ba
\label{eq:spin1}
\Sp &=& \frac{\omega}{\sqrt{\kappa^2-\J^2}}\,\frac{1}{2\,\eta\,\sqrt\eta}\,{}_2F_1\left(\frac{1}{2}, \frac{3}{2}, 2, -\frac{1}{\eta}\right), \\
\label{eq:energy1}
\E &=& \frac{\kappa}{\sqrt{\kappa^2-\J^2}}\,\frac{1}{\sqrt\eta}\,{}_2F_1\left(-\frac{1}{2}, \frac{1}{2}, 1, -\frac{1}{\eta}\right),  
\ea
Where $\kappa$ and $\omega$ are parameters of the classical solution.
From \refeq{spin1}, we can write $\eta = \eta(\Sp, \J)$ where $\J$ is treated as an expansion parameter
\be
\eta(\Sp, \J) = \eta^{(0)}(\Sp) + \eta^{(2)}(\Sp)\,\J^2 + \eta^{(4)}(\Sp)\,\J^4 + \cdots~.
\ee
This is the slow string limit which is the relevant one for the comparison with gauge theory results about finite twist operators.
The explicit functions $\eta^{(2n)}(\Sp)$ are at order $1/\Sp^4$ (we define $\bar{\cal S} = 8\,\pi\,{\cal S}$)
\ba
\label{eq:eta}
\eta^{(0)}(\Sp) &=& \frac{16}{\bar{\cal S}}+64 \left(3-\log \bar{\cal S}\right) \frac{1}{\bar{\cal S}^2}+64 \left(4 \log ^2\bar{\cal S}-30 \log \bar{\cal S}+35\right) \frac{1}{\bar{\cal S}^3}
\sxx +512 \left(-2
   \log ^3\bar{\cal S}+26 \log ^2\bar{\cal S}-75 \log \bar{\cal S}+52\right) \frac{1}{\bar{\cal S}^4}
+\dots~, \\
\eta^{(2)}(\Sp) &=& \frac{8\pi^2}{\bar\Sp\,\log^2\bar{\Sp}} + \frac{64\pi^2}{\bar{\Sp}^2}\left(
-\frac{1}{\log\bar{\Sp}}+\frac{3}{2\,\log^2\bar{\Sp}}+\frac{1}{\log^3\bar{\Sp}}
\right) +
\sxx +\frac{512\pi^2}{\bar{\Sp}^3} \left(\frac{3}{4}-\frac{27}{8 \log\bar\Sp}+\frac{17}{16 \log^2\bar\Sp}+\frac{7}{4 \log^3\bar\Sp}+\frac{3}{4 \log^4\bar\Sp}\right) +
\sxx +\frac{4096\pi^2}{\bar\Sp^4} \left(
-\frac{\log \bar{\Sp}}{2}+\frac{103}{24}-\frac{7}{\log \bar{\Sp}}-\frac{5}{8 \log ^2\bar{\Sp}}+\right.
\sxx\left. \frac{41}{24 \log ^3\bar{\Sp}}+\frac{3}{2 \log ^4\bar{\Sp}}+\frac{1}{2 \log ^5\bar{\Sp}}
\right) + \cdots~, \\
\eta^{(4)}(\Sp) &=& \frac{2 \pi ^4 \log^4\bar\Sp (4 \log\bar\Sp-1)}{\bar\Sp}
\sxx +16 \pi ^4 \log^4\bar\Sp \left(10 \log^2\bar\Sp-2 \log\bar\Sp-3\right) \,\frac{1}{\bar\Sp^2}
\sxx +8 \pi ^4 \log^2\bar\Sp \left(240 \log^5\bar\Sp-156 \log^3\bar\Sp-67 \log^2\bar\Sp+14 \log\bar\Sp+12\right)\, \frac{1}{\bar\Sp^3}
\sxx +\frac{128}{3}
   \pi ^4 \log\bar\Sp \left(420 \log^7\bar\Sp+120 \log^6\bar\Sp-400 \log^5\bar\Sp\right.
\sxx\left. -346 \log^4\bar\Sp-21 \log^3\bar\Sp+130 \log^2\bar\Sp+70 \log-24\right) \,\frac{1}{\bar\Sp^4} + \cdots~.
\ea
The quantum contribution to the energy is 
\be
\Delta = \E -\Sp  = \Delta^{(0)}(\Sp) + \Delta^{(2)}(\Sp)\,\J^2 + \Delta^{(4)}(\Sp)\,\J^4 + \cdots~.
\ee
and is obtained by replacing the above expansions in \refeq{energy1}. Again, we list the first functions $\Delta^{(2n)}(\Sp)$
\ba
\label{eq:delta}
\Delta^{(0)}({\cal S}) &=& \frac{\log\bar{\Sp}-1}{\pi} + \frac{4}{\pi}(\log\bar{\Sp}-1)\frac{1}{\bar{\Sp}}-\frac{4}{\pi}(2\log^2\bar{\Sp}-9\log\bar{\Sp}+5)\frac{1}{\bar{\Sp}^2} 
\sxx + 
\frac{32}{3\pi}(2\log^3\bar{\Sp}-18\log^2\bar{\Sp}+33\log\bar{\Sp}-14)\frac{1}{\bar{\Sp}^3}
\sxx -\frac{2}{3\pi}(96\log^4\bar{\Sp}-1376\log^3\bar{\Sp}+4896\log^2\bar{\Sp}-5622 \log\bar{\Sp}+1919)\frac{1}{\bar{\Sp}^4} + \cdots, \\
\Delta^{(2)}(\Sp) &=& \frac{\pi}{2\,\log\bar{\Sp}} + \frac{2\pi}{\bar{\Sp}\log^2\bar{\Sp}} + \frac{2\pi}{\bar{\Sp}^2}
\left(-2-\frac{3}{\log\bar{\Sp}}+\frac{2}{\log^2\bar{\Sp}}+\frac{4}{\log^3\bar{\Sp}}\right) +
\sxx +32\,\pi\left(\frac{2 \log \bar\Sp}{3}-1-\frac{2}{\log \bar\Sp}-\frac{2}{3 \log ^2\bar\Sp}+\frac{1}{\log ^3\bar\Sp}+\frac{1}{\log ^4\bar\Sp}\right)\,\frac{1}{\bar\Sp^3}
\sxx -96\pi\left(\log ^2\bar\Sp-\frac{16 \log \bar\Sp}{3}+\frac{11}{9}+\frac{75}{16 \log \bar\Sp}\right.
\sxx\left. +\frac{123}{32 \log ^2\bar\Sp}+\frac{4}{9 \log ^3\bar\Sp}-\frac{2}{\log ^4\bar\Sp}-\frac{4}{3 \log ^5\bar\Sp}\right)\,\frac{1}{\bar\Sp^4} + \cdots, \\
\Delta^{(4)}(\Sp) &=& \frac{\pi^3}{8}\left(\frac{1}{\log ^4\bar\Sp}-\frac{1}{\log ^3\bar\Sp}\right) 
\sxx +\pi^3\left(\frac{2}{\log ^5\bar\Sp}-\frac{3}{\log ^4\bar\Sp}\right)\,\frac{1}{\bar\Sp}
\sxx + \pi^3\left( \frac{1}{\log ^2\bar\Sp}+\frac{15}{2 \log ^3\bar\Sp}-\frac{11}{2 \log ^4\bar\Sp}-\frac{32}{\log ^5\bar\Sp}+\frac{20}{\log ^6\bar\Sp}\right)\,
\frac{1}{\bar\Sp^2}
\sxx +8\,\pi^3\left(\frac{2}{3 \log ^2\bar\Sp}+\frac{10}{\log ^3\bar\Sp}+\frac{13}{\log ^4\bar\Sp}-\frac{50}{3 \log ^5\bar\Sp}-\frac{30}{\log ^6\bar\Sp}+\frac{20}{\log ^7\bar\Sp}
\right)\,\frac{1}{\bar\Sp^3}
\sxx +\pi^3\left(-24-\frac{120}{\log \bar\Sp}-\frac{232}{\log ^2\bar\Sp}+\frac{2665}{6 \log ^3\bar\Sp}+\frac{5697}{4 \log ^4\bar\Sp}\right.
\sxx \left. +\frac{719}{\log ^5\bar\Sp}-\frac{5120}{3 \log ^6\bar\Sp}-\frac{1440}{\log
   ^7\bar\Sp}+\frac{1120}{\log ^8\bar\Sp}\right)\,\frac{1}{\bar\Sp^4} + \cdots.
\ea
The quantity $\Delta$  is reciprocity respecting in the following sense. We first define the function $f$ by 
\be
\Delta(\Sp) = {\cal E}({\cal S})-{\cal S} = f\left({\cal S} + \frac{1}{2}\,{\cal E}({\cal S})\right)
\ee
This is a good definition at large ${\cal S}$ since ${\cal E}({\cal S})\sim \log{\cal S}$ 
and the argument of $f$ can be treated perturbatively.\\ 


Applying the Lagrange-B\"urmann formula, we find (Eq.(3.7) of \cite{Basso:2006nk})
\be
f({\cal S}) = \sum_{k=1}^\infty\frac{1}{k!}\left(-\frac{1}{2}\frac{d}{d{\cal S}}\right)^{k-1}[\Delta({\cal S})]^k.
\ee
Again we expand in powers of $\J$, 
\be
f(\Sp)  = f^{(0)}(\Sp) + f^{(2)}(\Sp)\,\J^2 + f^{(4)}(\Sp)\,\J^4 + \cdots~.
\ee
and a straightforward calculation gives
\ba
\label{eq:f0}
f^{(0)}({\cal S}) &=& \frac{\log\bar{\Sp}-1}{\pi} + \mbox{\fbox{$\displaystyle 0\cdot\frac{1}{\bar{\Sp}}$}}
\sxx + \frac{4}{\pi}(\log\bar{\Sp}+1)\,\frac{1}{\bar{\Sp}^2} + \mbox{\fbox{$\displaystyle 0\cdot\frac{1}{\bar{\Sp}^3}$}}
\sxx - \frac{2}{\pi}(16\log^2\bar{\Sp}+14\log\bar{\Sp}+5)\,\frac{1}{\bar{\Sp}^4} + \mbox{\fbox{$\displaystyle 0\cdot\frac{1}{\bar{\Sp}^5}$}}
+\cdots~, \\
f^{(2)}({\cal S}) &=& \frac{\pi}{2\,\log\bar{\Sp}} +
\mbox{\fbox{$\displaystyle 0\cdot\frac{1}{\bar{\Sp}}$}}
\sxx -\frac{6\pi}{\log\bar{\Sp}}\,\frac{1}{\bar{\Sp}^2} +
\mbox{\fbox{$\displaystyle 0\cdot\frac{1}{\bar{\Sp}^3}$}}
\sxx + \frac{\pi (-1+30\,\log\bar{\Sp}+80\,\log\bar{\Sp}^2)}{\log\bar{\Sp}^2}\, \frac{1}{\bar{\Sp}^4}+
\mbox{\fbox{$\displaystyle 0\cdot\frac{1}{\bar{\Sp}^5}$}}+ \cdots~, \\
f^{(4)}({\cal S}) &=& -\frac{\pi ^3 (\log \bar\Sp-1)}{8 \log ^4\bar\Sp}+
\mbox{\fbox{$\displaystyle 0\cdot\frac{1}{\bar{\Sp}}$}}
\sxx +\frac{\pi ^3 (7 \log \bar\Sp-5)}{2 \log ^4\bar\Sp}\,\frac{1}{\bar\Sp^2}+
\mbox{\fbox{$\displaystyle 0\cdot\frac{1}{\bar{\Sp}^3}$}}
\sxx -\frac{\pi ^3 \left(304 \log ^3\bar\Sp+26 \log ^2\bar\Sp-81 \log
   \bar\Sp+4\right)}{4 \log ^5\bar\Sp }\,\frac{1}{\bar\Sp ^4 }+
\mbox{\fbox{$\displaystyle 0\cdot\frac{1}{\bar{\Sp}^5}$}}+\cdots~.
\ea
Reciprocity is the absence of inverse odd powers of $\Sp$ in the above expansions (the terms inside boxes).

\subsection{Higher conserved charges and their reciprocity}

Let us consider the $(J_1, J_2)$ string and its higher charges appearing
in~\cite{Arutyunov:2003rg}, Section 3.3  and reviewed
in App.~(\ref{app:as}).
In~\cite{Arutyunov:2003rg} Arutyunov and Staudacher analyzed the matching
between the conserved charged for the closed  $\mathfrak{su}(2)$ sector; in
the strong coupling regimes they constructed explicitly the higher charges 
by using  the B\"acklund transformations in the 
integrable classical string $\sigma$-model.
 The first one beyond the energy is 
\be
{\cal E}_4 = -\frac{16}{\pi^2\,{\cal E}_2}\,Z_1(t) + \frac{32}{\pi^4\,{\cal E}_2^3}\, Z_2(t), 
\ee
where
\ba
Z_1(t) &=& \mathbb{K}(t)[\mathbb{E}(t)+(t-1)\mathbb{K}(t)], \\
Z_2(t) &=& t(t-1) \mathbb{K}(t)^4,
\ea
and $t$ is a modular parameter. \\

The two $\sigma$-models describing string on  $AdS_3 \times S^1$ and $R
\times S^3$ are simply related  by analytic continuation of coordinates,
and the  conserved charges  defined  in ~\cite{Arutyunov:2003rg}
are as well  expressed in terms of $\sigma$-model coordinates;  equations of motion, their solutions 
and the charges are mapped by analytic continuation from one $\sigma$-model
into another\footnote{We thank A. A. Tseytlin for useful discussions on this point}.
Upon analytic continuation to the $(S, J)$ string we know that 
\be
t\to -1/\eta, \qquad {\cal E}_2\to J.
\ee
In analogy to the case of the energy, we propose to identify the coefficients $Z_k(t)$ of the various powers of $1/J$ as non-trivial functions of the modular
parameter $\eta$ which are reciprocity respecting~\footnote{Notice, that it is non trivial to relate this quantities 
to the weak coupling charges $q_r$. These have certainly a well-defined strong coupling limit as discussed in~\cite{Belitsky:2007kf}, and an investigation of 
their reciprocity properties is an interesting problem.}. Now, reciprocity must be tested on the functions $f_k$ defined by 
\be
\label{eq:functional}
Z_k({\cal S}) = f_k\left({\cal S} + \frac{1}{2}{\cal E}({\cal S})\right),
\ee
(where we have defined $Z_k({\cal S}) \equiv Z_k(-1/\eta({\cal S}))$).\\

The Lagrange-B\"urmann formula takes now the following form (we omit for simplicity the index $k$)
\be
f({\cal S}) = \sum_{k=0}^\infty\frac{1}{k!}\left(\frac{d}{d{\cal S}}\right)^{k-1}\left[
\left(-\frac{\Delta({\cal S})}{2}\right)^k\,Z'({\cal S})\right] = Z({\cal S})-\frac{1}{2}\,\Delta({\cal S})\,Z'({\cal S}) + \cdots
\ee
Notice that $f$ depends linearly on $Z$. Thus, linear combinations of reciprocity respecting quantities are reciprocity respecting. This linearity
is due to the fact that $Z$ does not appear in the argument of $f$ in the functional relation \refeq{functional}.
Since $\eta = \eta(\Sp, \J)$ has a non trivial $\J$ dependence, we have again an expansion 
\be
f_k(\Sp)  = f^{(0)}_k(\Sp) + f^{(2)}_k(\Sp)\,\J^2 + f^{(4)}_k(\Sp)\,\J^4 + \cdots~.
\ee
Working out the 0-th order correction for $Z_1$ and $Z_2$ we find the result
\ba
\label{eq:fz1}
f_1^{(0)} &=& -\frac{1}{4} \left(\log \bar{\Sp}-2\right) \log \bar{\Sp} + 
\mbox{\fbox{$\displaystyle 0\cdot\frac{1}{\bar{\Sp}}$}}
\sxx +2 \left(2-3 \log \bar{\Sp}\right) \log \bar{\Sp} \,\frac{1}{\bar{\Sp}^2} +
\mbox{\fbox{$\displaystyle 0\cdot\frac{1}{\bar{\Sp}^3}$}}
\sxx +\left(80 \log
   ^3\bar{\Sp}-118 \log ^2\bar{\Sp}+23 \log \bar{\Sp}+1\right) \,\frac{1}{\bar{\Sp}^4}+
\mbox{\fbox{$\displaystyle 0\cdot\frac{1}{\bar{\Sp}^5}$}}+\cdots, \\
f_2^{(0)} &=& \frac{1}{16} \log ^4\bar{\Sp}+
\mbox{\fbox{$\displaystyle 0\cdot\frac{1}{\bar{\Sp}}$}}
\sxx +\log ^4\bar{\Sp} \,\frac{1}{\bar{\Sp}^2} +
\mbox{\fbox{$\displaystyle 0\cdot\frac{1}{\bar{\Sp^3}}$}}
\sxx -\frac{1}{2} \left(\log ^3\bar{\Sp} \left(16 \log ^2\bar{\Sp}-22 \log
   \bar{\Sp}-1\right)\right) \,\frac{1}{\bar{\Sp}^4}+
\mbox{\fbox{$\displaystyle 0\cdot\frac{1}{\bar{\Sp^5}}$}}.
\ea
In both cases, there is parity invariance. Going to the next charge ${\cal E}_6$, we find the new structures
\ba
Z_3 &=& \mathbb{K}(t)^3[(8t-4)\mathbb{E}(t)+(t-1)(15t-4)\mathbb{K}(t)], \\
Z_4 &=& t(t-1)\mathbb{K}(t)^5[\mathbb{E}(t)+(3t-2)\mathbb{K}], \\
Z_5 &=& t^2(t-1)^2 \mathbb{K}(t)^8 = Z_2(t)^2.
\ea
The calculation of $f_k$ gives 
\ba
f_3^{(0)} &=& \frac{1}{16} \log ^3\bar{\Sp} \left(15 \log \bar{\Sp}-16\right) +
\mbox{\fbox{$\displaystyle 0\cdot\frac{1}{\bar{\Sp}}$}}
\sxx +3 \log ^3\bar{\Sp} \left(13 \log \bar{\Sp}-16\right) \frac{1}{\bar{\Sp}^2}+
\mbox{\fbox{$\displaystyle 0\cdot\frac{1}{\bar{\Sp^3}}$}}
\sxx -\frac{3}{2} \left(\log
   ^2\bar{\Sp} \left(336 \log ^3\bar{\Sp}-1102 \log ^2\bar{\Sp}+619 \log \bar{\Sp}+4\right)\right)
   \frac{1}{\bar{\Sp}^4}+
\mbox{\fbox{$\displaystyle 0\cdot\frac{1}{\bar{\Sp^5}}$}}, \\
f_4^{(0)} &=& \frac{1}{64} \left(2-3 \log \bar{\Sp}\right) \log ^5\bar{\Sp} +
\mbox{\fbox{$\displaystyle 0\cdot\frac{1}{\bar{\Sp}}$}}
\sxx +\frac{1}{8} \left(6-19 \log \bar{\Sp}\right) \log ^5\bar{\Sp} \frac{1}{\bar{\Sp}^2} +
\mbox{\fbox{$\displaystyle 0\cdot\frac{1}{\bar{\Sp^3}}$}}
\sxx +\frac{1}{16} \log
   ^4\bar{\Sp} \left(464 \log ^3\bar{\Sp}-1014 \log ^2\bar{\Sp}+167 \log \bar{\Sp}+5\right) 
\frac{1}{\bar{\Sp}^4}+
\mbox{\fbox{$\displaystyle 0\cdot\frac{1}{\bar{\Sp^5}}$}} , \\
f_5^{(0)} &=& \frac{1}{256} \log ^8\bar{\Sp} + 
\mbox{\fbox{$\displaystyle 0\cdot\frac{1}{\bar{\Sp}}$}}
\sxx +\frac{1}{8} \log ^8\bar{\Sp} \frac{1}{\bar{\Sp}^2}+
\mbox{\fbox{$\displaystyle 0\cdot\frac{1}{\bar{\Sp^3}}$}}
\sxx -\frac{1}{16} \left(\log ^7\bar{\Sp} \left(16 \log ^2\bar{\Sp}-38 \log
   \bar{\Sp}-1\right)\right) \frac{1}{\bar{\Sp}^4}+
\mbox{\fbox{$\displaystyle 0\cdot\frac{1}{\bar{\Sp^5}}$}}~.
\ea
Again, parity invariance is observed.\\

The next corrections in $\J^2$ are also parity invariant, precisely as happened in the case of the energy. To give an example, the first two corrections 
the the function $f_1$ associated with the structure $Z_1(t)$ are 
\ba
f_1^{(2)} &=& \frac{\pi^2(-1+\log\bar{\Sp})}{4\log\bar{\Sp}^2} +
\mbox{\fbox{$\displaystyle 0\cdot\frac{1}{\bar{\Sp}}$}}
\sxx + 2\pi^2\,\frac{1}{\bar{\Sp}^2} +
\mbox{\fbox{$\displaystyle 0\cdot\frac{1}{\bar{\Sp}^3}$}} 
\sxx + \frac{\pi^2 (2 + 7\log\bar{\Sp}+12\log\bar{\Sp}^2+184\log\bar{\Sp}^3-192\log\bar{\Sp}^4)}{2\log\bar{\Sp}^3}\frac{1}{\bar{\Sp}^4} +
\mbox{\fbox{$\displaystyle 0\cdot\frac{1}{\bar{\Sp}^5}$}}+ \cdots~, \\
f_1^{(4)} &=& 
-\frac{\pi ^4 \left(2 \log ^2\bar\Sp-5 \log \bar\Sp+4\right)}{16 \log ^5\bar\Sp} + 
\mbox{\fbox{$\displaystyle 0\cdot\frac{1}{\bar{\Sp}}$}}
\sxx
\frac{\pi ^4 \left(2 \log ^2\bar\Sp-7 \log \bar\Sp+4\right)}{2  \log ^5\bar\Sp}\,\frac{1}{\bar\Sp^2} + 
\mbox{\fbox{$\displaystyle 0\cdot\frac{1}{\bar{\Sp}^3}$}} 
\sxx \!\!\!\!\!\!
+\frac{\pi ^4 \left(64 \log ^5\bar\Sp+88 \log ^4\bar\Sp+40 \log ^3\bar\Sp-23 \log ^2\bar\Sp+6 \log \bar\Sp+10\right)}{4\log ^6\bar\Sp}\,\frac{1}{ \bar\Sp^4 }+
\mbox{\fbox{$\displaystyle 0\cdot\frac{1}{\bar{\Sp}^5}$}}+ \cdots.\nonumber\\
\ea

We tested in this way all the structures  appearing in App.~(\ref{app:as}), always finding that the reciprocity condition in
satisfied.

\section*{Acknowledgments}
We thank  A. A. Tseytlin and G. P. Korchemsky for valuable discussions during this work and 
for  kind encouragement. 

\newpage
\appendix

\section{The Arutyunov-Staudacher conserved charges for the $(J_1, J_2)$ string}
\label{app:as}

In this Appendix, we review the results of~\cite{Arutyunov:2003rg} and give a list of explicit results which 
are needed in the strong coupling analysis of reciprocity. The conserved charges $\E_{2n}$ for the $(J_1, J_2)$ string
can be obtained from the expansion  
\be
\E(\gamma) = \sum_{n\ge 2} \E_n\,\gamma^2,
\ee
where~\footnote{$\Pi(u\,|\,k)$ is the complete elliptic integral of III kind.}
\be
\E(\gamma) = \frac{4\,\gamma^3}{\pi\,(1+\gamma^2)}\sqrt\frac{(1-z)(1-t\,z)}{z}\,\Pi(t\,z | z),
\ee
and $z=z(\gamma, t)$ is obtained from the power series expansion in $\gamma^2$ of 
\be
1-\frac{\omega_1^2}{\omega_2^2-\omega_1^2}\,\frac{z}{z-1}-\left(\frac{1-\gamma^2}{1+\gamma^2}\right)^2\,\frac{1}{1-t\,z} = 0,
\ee
with 
\ba
\omega_1^2 &=& \E_2^2-\frac{4}{\pi^2}\,t\,\mathbb{K}(t)^2, \\
\omega_2^2 &=& \E_2^2-\frac{4}{\pi^2}\,(t-1)\,\mathbb{K}(t)^2,
\ea
(of course one must choose the branch $z(0, t)=0$). 
Using the expansion 
\ba
\Pi(\varepsilon | q) &=& \mathbb{K}(q)+\frac{(\mathbb{K}(q)-\mathbb{E}(q)) \varepsilon }{q}+\frac{((q+2) \mathbb{K}(q)-2 (q+1) \mathbb{E}(q)) \varepsilon ^2}{3 q^2}
\sxx +\frac{\left(\left(4 q^2+3 q+8\right) \mathbb{K}(q)-\left(8 q^2+7 q+8\right) \mathbb{E}(q)\right) \varepsilon ^3}{15 q^3}+\cdots, 
\ea
we find the general structure ($\E_2$ is the energy of the $(J_1, J_2)$ string to be analytically continued to the angular momentum $\J$ of the $(S, J)$ string)
\be
\E_{2n} = \sum_{p=1}^{n-2}\frac{Z_{2n, p}(t)}{\pi^{2p}\,\E_2^{2p-1}},
\ee
where a list of the $Z$ appearing in the first 10 charges is 
\ba
Z_{4,1} &=& -16 \mathbb{K}(t) (\mathbb{E}+(t-1) \mathbb{K}(t)), \\
Z_{4,2} &=& 32 (t-1) t \mathbb{K}(t)^4, \\
Z_{6,1} &=& 32 \mathbb{K}(t) (\mathbb{E}+(t-1) \mathbb{K}(t)), \\
Z_{6,2} &=& -\frac{64}{3} \mathbb{K}(t)^3 \left((8 t-4) \mathbb{E}+\left(15 t^2-19 t+4\right) \mathbb{K}(t)\right), \\
Z_{6,3} &=& 512 (t-1) t \mathbb{K}(t)^5 (\mathbb{E}+(3 t-2) \mathbb{K}(t)), \\
Z_{6,4} &=& -2560 (t-1)^2 t^2 \mathbb{K}(t)^8, \\
Z_{8,1} &=& -48 \mathbb{K}(t) (\mathbb{E}+(t-1) \mathbb{K}(t)), \\
Z_{8,2} &=& \frac{32}{3} \mathbb{K}(t)^3 \left(32 (2 t-1) \mathbb{E}+\left(105 t^2-137 t+32\right) \mathbb{K}(t)\right), \\
Z_{8,3} &=& -\frac{2048}{5} \mathbb{K}(t)^5 \left(\left(17 t^2-17 t+2\right) \mathbb{E}+\left(35 t^3-61 t^2+28 t-2\right) \mathbb{K}(t)\right),\\
Z_{8,4} &=& 2048 (t-1) t \mathbb{K}(t)^7 \left(8 (2 t-1) \mathbb{E}+\left(45 t^2-53 t+12\right) \mathbb{K}(t)\right), \\
Z_{8,5} &=& -57344 (t-1)^2 t^2 \mathbb{K}(t)^9 (\mathbb{E}+(5 t-3) \mathbb{K}(t)), \\
Z_{8,6} &=& 344064 (t-1)^3 t^3 \mathbb{K}(t)^{12}, \\
Z_{10,1} &=& 64 \mathbb{K}(t) (\mathbb{E}+(t-1) \mathbb{K}(t)), \\
Z_{10,2} &=& -\frac{128}{3} \mathbb{K}(t)^3 \left(20 (2 t-1) \mathbb{E}+\left(63 t^2-83 t+20\right) \mathbb{K}(t)\right),\\
Z_{10,3} &=& \frac{1024}{5} \mathbb{K}(t)^5 \left(\left(169 t^2-169 t+24\right) \mathbb{E}+\right. 
\sxx \left. \left(315 t^3-557 t^2+266 t-24\right) \mathbb{K}(t)\right),\\
Z_{10,4} &=& -\frac{1024}{7} \mathbb{K}(t)^7 \left(32 \left(82 t^3-123 t^2+45 t-2\right) \mathbb{E}+\right.
\sxx\left. \left(5775 t^4-12862 t^3+9055 t^2-2032 t+64\right) \mathbb{K}(t)\right),\\
Z_{10,5} &=& 16384 (t-1) t \mathbb{K}(t)^9 \left(3 \left(47 t^2-47 t+8\right) \mathbb{E}+\right.
\sxx \left. \left(385 t^3-648 t^2+303 t-32\right) \mathbb{K}(t)\right),\\
Z_{10,6} &=& -294912 (t-1)^2 t^2 \mathbb{K}(t)^{11} \left(12 (2 t-1) \mathbb{E}+\left(91 t^2-103 t+24\right) \mathbb{K}(t)\right),\\
Z_{10,7} &=& 8650752 (t-1)^3 t^3 \mathbb{K}(t)^{13} (\mathbb{E}+(7 t-4) \mathbb{K}(t)),\\
Z_{10,8} &=& -56229888 (t-1)^4 t^4 \mathbb{K}(t)^{16}.
\ea


\end{document}